\documentclass[11pt,twocolumn]{article}

\usepackage[utf8]{inputenc}
\usepackage{amsmath, amssymb}
\usepackage{graphicx}
\usepackage{cite}          
\usepackage{authblk}       
\usepackage{hyperref}      
\usepackage{cuted}
\usepackage{mathtools}
\usepackage{natbib}
\usepackage{physics}
\usepackage{tikz}
\usepackage{braket}

\title{$S_8-H_0$ tension in a SI-ULDM scenario}
\author[1]{Jessica N. López-Sánchez }

\affil[1]{\small CEICO---FZU, Institute of Physics of the Czech Academy of Sciences, Na Slovance 1999/2, 182 00 Prague, Czech Republic}
\date{\today}

\begin{document}

\maketitle
\begin{abstract}
We study the cosmological impact of a transient self--interaction phase in ultra--light dark matter (ULDM), focusing on its simultaneous effects on the sound horizon and the late--time growth of structure. In the presence of a quartic self--interaction, the scalar field undergoes a short--lived radiation--like phase before evolving into matter--like behaviour, inducing a localized modification of the expansion history at early times. We develop a perturbative and model--independent framework in which the self--interaction energy density is parametrized as a localized contribution to the total energy budget. Within this approach, the responses of the sound horizon and the linear growth factor can be expressed as weighted integrals over cosmic time, with distinct kernels encoding the temporal sensitivity of each observable. This structure leads to a simple analytic relation linking the corresponding early-- and late--time responses, and naturally predicts correlated shifts in $H_0$ and $S_8$ whose sign and magnitude depend on the timing of the self--interaction episode.

Our results show that a single transient modification of the expansion history can interpolate between early--time effects on the sound horizon and late--time suppression of structure growth within a unified physical framework, providing an analytical understanding of their joint response.

\end{abstract}

\section{Introduction}

The standard cosmological model ($\Lambda$CDM) provides an excellent description of a wide range of observations, from the Cosmic Microwave Background (CMB) to large--scale structure \cite{aghanim2020planck,ross2017clustering,weinberg2013observational}. Nevertheless, some tensions have emerged between early-- and late--time cosmological parameters. In particular, there is a discrepancy between the value of the Hubble constant inferred from the CMB within $\Lambda$CDM and local distance measurements, currently at the level of $\sim5\sigma$ \cite{riess2019large,riess2022comprehensive,verde2019tensions}. Similarly, measurements of the amplitude of matter fluctuations indicate that weak lensing and large--scale structure surveys favor lower values of $S_8$ than those inferred from Planck assuming $\Lambda$CDM \cite{heymans2021kids,abbott2022dark}. Although these discrepancies do not yet constitute a breakdown of $\Lambda$CDM, they may point to the presence of new physical ingredients beyond the minimal model.

Proposed solutions to these tensions can be broadly classified into two categories \cite{verde2019tensions, knox2020hubble, di2021realm}. Late--time modifications of dark energy or gravity can suppress the growth of structure and thus alleviate the $S_8$ tension, but they typically leave the sound horizon unchanged and therefore do not address the discrepancy in $H_0$ \cite{linder2020limited, knox2020hubble, efstathiou2021h}. By contrast, early--time solutions modify the expansion history prior to recombination, reducing the sound horizon and increasing the CMB--inferred value of $H_0$ \cite{poulin2019early, ivanov2020constraining}. A key challenge for this class of models is to produce such early--time effects while maintaining a controlled and predictive impact on the late--time growth of structure.

Self--Interacting Ultra--Light Dark Matter (SI--ULDM) provides a well--motivated framework in which early--time modifications can arise naturally \cite{hu2000fuzzy, hui2017ultralight}. In the presence of self--interactions, an ultra--light scalar field can undergo a transient phase during which its effective equation of state differs from that of Cold Dark Matter (CDM). In particular, a quartic self--interaction induces radiation--like behaviour at early times, followed by a transition to a matter--like phase once the quadratic mass term becomes dominant \cite{amin2015nonperturbative, chavanis2011mass, shapiro2022cosmological}.

ULDM has been extensively studied at both linear and non--linear levels \cite{hlozek2015search, marsh2016axion, shapiro2022cosmological, lopez2025core, mocz2023cosmological}. In particular, previous studies of Early Eark Energy (EDE) and SI--ULDM scenarios have assessed the phenomenological viability of early--time modifications to the expansion history using numerical methods and model--dependent implementations, often relying on full Boltzmann solvers and data--driven parameter constraints \cite{poulin2019early, santos2017constraining, hill2020early, chavanis2011mass}. However, a unified analytical description connecting the impact of a single localized episode on the sound horizon and on the late--time growth of structure has not yet been fully developed, leaving open the question of how correlated modifications in $H_0$ and $S_8$ arise.

In this work, we develop a perturbative and model--independent framework to quantify the cosmological impact of a transient quartic phase of SI--ULDM. By parametrizing the self--interaction energy density as a localized contribution to the total energy budget, we show that its effects on the sound horizon and on the linear growth of structure can be expressed as weighted integrals over cosmic time. This leads to a simple analytic relation linking the early-- and late--time responses, distinguishing this scenario from purely late--time modifications of the cosmological model discussed in the literature.

\section{Background}
\label{sec:quartic_transition}

We consider an ultra--light real scalar field $\phi$ with mass $m$ and a quartic self--interaction, described by the Lagrangian \cite{hu2000fuzzy}
\begin{equation}
\mathcal{L}
=\frac{1}{2}\,\partial_\mu\phi\,\partial^\mu\phi
-\frac{1}{2}m^2\phi^2
-\frac{\lambda}{4}\phi^4 ,
\end{equation}
where $\lambda$ is the self--interaction coupling. In the non--relativistic limit, it can be related to an effective two--body scattering length.

At early times, when the field amplitude is large and $\lambda\langle\phi^2\rangle \gg m^2$, the quartic term dominates the potential. In this regime, the field oscillates in a quartic potential and behaves as an additional relativistic  component,with an averaged equation of state $w\simeq 1/3$ and energy density $\rho_{\rm SI}\propto a^{-4}$ \cite{turner1983coherent, preskill1983cosmology}. As the universe expands, the amplitude of $\phi$ redshifts and the quadratic mass term eventually becomes dominant, $m^2 \gg \lambda\langle\phi^2\rangle$, leading to $w\simeq 0$ and $\rho_m\propto a^{-3}$, so that the field becomes indistinguishable from CDM \cite{chavanis2011mass, li2014cosmological}.

The transition between these two regimes in the scalar--field dynamics defines a characteristic scale factor $a_\star$ (or redshift $z_\star$), at which the quartic self--interaction and the quadratic mass term contribute comparably to the energy density,
\begin{equation}\label{eq:transition}
   \rho_{\rm SI}
   \equiv \frac{\lambda}{4}\langle\phi^4\rangle
   \;\sim\;
   \rho_m
   \equiv \frac{m^2}{2}\langle\phi^2\rangle .
\end{equation}

\subsection{Phenomenological parametrization}

Following the EDE literature \cite{poulin2019early, hill2020early}, we model the fractional contribution of the quartic component as a smooth ``bump'' in $\ln a$,
\begin{equation}
f_{\rm SI}(a) \;\equiv\; \frac{\rho_{\rm SI}(a)}{\rho_{\rm tot}(a)}
= f_{\rm pk}\,
\exp\!\left[-\frac{(\ln(a/a_\star))^2}{2\Delta^2}\right].
\label{eq:fSI}
\end{equation}
Here $f_{\rm pk}$ is the peak fractional contribution relative to the total energy density, $a_\star$ denotes the scale factor around which the quartic--quadratic transition occurs (i.e.~the
epoch characterized by Eq.~\ref{eq:transition}), and $\Delta$ sets the width of the bump in $\ln a$. Throughout this work we assume a subdominant self--interaction component,
$f_{\rm SI}(a)\le f_{\rm pk}\ll 1$, for all $a$. This parametrization has three key advantages: (i) it vanishes both at early and late times, (ii) it peaks at the expected transition epoch, and (iii) it captures the main impact of the quartic phase on the expansion history with only two effective parameters, $(f_{\rm pk},\Delta)$, which can later be related to the underlying microphysical quantities $(m,\lambda)$.

For $a \lesssim a_\star$, when $f_{\rm SI}(a)$ is non--negligible and the field behaves as a radiation--like component, the expansion rate is temporarily enhanced relative to $\Lambda$CDM. This localized modification of $H(a)$ is the source of the changes to the sound horizon and, more indirectly, to the late--time growth of structure discussed in the following sections.

\section{Impact on the sound horizon}

In order to describe the impact of the self--interaction phase on cosmological observables, we begin by considering the sound horizon. This provides the most direct probe of early--time modifications of the expansion history, as it depends solely on the background expansion. At recombination, the sound horizon is given by
\begin{equation}
r_s(a_{\rm rec})
=\int_0^{a_{\rm rec}}
\frac{c_s(a)}{a^2 H(a)}\,da.
\end{equation}
This expression sets the physical scale of the acoustic peaks in the CMB and thus plays a central role in the inference of the Hubble constant within $\Lambda$CDM \cite{hu1996small, eisenstein1998baryonic}. A temporary increase in the expansion rate prior to recombination shortens the distance travelled by acoustic waves, resulting in a smaller sound horizon and a correspondingly larger CMB--inferred value of $H_0$.

The sound speed of the tightly coupled photon--baryon plasma is given by \cite{hu2002cosmic}
\begin{equation}
c_s(a)=\frac{1}{\sqrt{3(1+R(a))}},
\qquad
R(a)\equiv\frac{3\rho_b}{4\rho_\gamma}.
\end{equation}
Here, $\rho_b$ and $\rho_\gamma$ denote the energy densities of baryons and photons, respectively. Note that only these components contribute to $c_s$, since they form the oscillating fluid. All other components, including neutrinos, CDM, and the scalar field, affect the sound horizon only through their contribution to the background expansion rate $H(a)$.

\subsection{Linear response to a small self--interaction component}
\label{sec:kernel_sound}

To first order in the small fraction $f_{\rm SI}\ll1$, the Hubble rate can be expanded as
\begin{equation}\label{eq:hubble}
H^2(a)=H^2_{\Lambda{\rm CDM}}(a)\,[1+f_{\rm SI}(a)],
\end{equation}
so that
\begin{equation}
\frac{1}{H(a)}
\simeq
\frac{1}{H_{\Lambda{\rm CDM}}(a)}
\left[1-\frac{1}{2}f_{\rm SI}(a)\right].
\end{equation}
Since the self--interaction affects only the background expansion, the sound speed $c_s$ remains unchanged. The fractional shift in the sound horizon is therefore controlled entirely by the overlap between $f_{\rm SI}(a)$ and the sound--horizon integrand,
\begin{equation}
\frac{\Delta r_s}{r_s}
\simeq
-\frac12
\frac{\displaystyle\int_0^{a_{\rm rec}}
\frac{c_s}{a^2H}\,f_{\rm SI}(a)\,da}
{\displaystyle\int_0^{a_{\rm rec}}\frac{c_s}{a^2H}\,da}
\equiv
-\frac12\,\langle f_{\rm SI}\rangle_{r_s}.
\label{eq:drs}
\end{equation}

Equation~(\ref{eq:drs}) shows that $\langle f_{\rm SI}\rangle_{r_s}$ is a weighted average of the self--interaction fraction, with weight given by the sound--horizon kernel $K(a)=c_s/(a^2H)$. Only values of $f_{\rm SI}(a)$ that overlap with the epoch in which this kernel is large contribute appreciably to $\Delta r_s$. In other words, the sound horizon is sensitive not to the absolute amplitude of the self--interaction bump, but to the portion that lies within the acoustic propagation window. This structure is analogous to that found in EDE scenarios, where the impact on $r_s$ is similarly controlled by the overlap between the EDE fraction and the sound--horizon kernel \cite{poulin2019early, hill2020early}.

\subsection{Sound-horizon kernel representation}

For analytic insight, it is convenient to rewrite the integrals in terms of $d\ln a$ and to define the weight
\begin{equation}\label{eq:sound_kernel}
\tilde K(a)\equiv \frac{c_s}{aH},
\end{equation}
which corresponds to the sound--horizon kernel expressed in logarithmic time. Therefore, equation (\ref{eq:drs}) can then be written as
\begin{equation}
\langle f_{\rm SI}\rangle_{r_s}
=\frac{\int d\ln a\,\tilde K(a)\,f_{\rm SI}(a)}
{\int d\ln a\,\tilde K(a)}.
\end{equation}
If $f_{\rm SI}(a)$ is given by the narrow Gaussian bump of Eq.~\eqref{eq:fSI}, and if the kernel $\tilde K(a)$ varies slowly across the width $\Delta$, one finds
\begin{equation}
\langle f_{\rm SI}\rangle_{r_s}
\simeq
f_{\rm pk}\,
\frac{\tilde K(a_\star)\,\sqrt{2\pi}\Delta}
{\int d\ln a\,\tilde K(a)}.
\end{equation}
Introducing the effective weight
\begin{equation}
\mathcal W_\star
\equiv
\frac{\tilde K(a_\star)\,\sqrt{2\pi}\Delta}
{\int d\ln a\,\tilde K(a)},
\end{equation}
we obtain the compact result
\begin{equation}\label{eq:delta_rs}
\frac{\Delta r_s}{r_s}
\simeq
-\frac12\,f_{\rm pk}\,\mathcal W_\star.
\end{equation}

This relation makes explicit how the position, width, and amplitude of the self--interaction bump determine its effective contribution to the sound horizon. Contributions centered near matter--radiation equality receive the largest weight, reflecting the fact that the sound--horizon kernel peaks around this epoch (see, e.g., \cite{knox2020hubble, smith2021early}). By contrast, bumps occurring much earlier or much later are strongly suppressed.

\subsection{Connection to $H_0$}

The CMB measures the angular acoustic scale $\theta_\star \equiv r_s/D_A$ with percent--level precision \cite{hu1996small, aghanim2020planck}. Therefore, a reduction in the sound horizon must be compensated by a corresponding decrease in the angular diameter distance to recombination in order to keep $\theta_\star$ fixed, which in turn leads to a larger inferred value of the Hubble constant.
Models that reduce the sound horizon, such as EDE scenarios, generically predict an increase in the CMB--inferred value of $H_0$. However, due to parameter degeneracies in CMB fits, this response is not one--to--one. As a result, the relation between $\Delta H_0/H_0$ and $\Delta r_s/r_s$ can be characterized by an effective proportionality factor $|\eta| \lesssim 1$ in linearized treatments (see e.g. \cite{poulin2019early, knox2020hubble,aylor2019sounds, efstathiou2021h}and references therein), that is:
\begin{equation}\label{eq:delta_h0}
\frac{\Delta H_0}{H_0}
\simeq \eta\,\frac{\Delta r_s}{r_s},
\qquad 
|\eta|\lesssim \mathcal{O}(1).
\end{equation}
Hence an earlier quartic phase of SI--ULDM, which suppresses $r_s$, naturally leads to a larger $H_0$ inferred from CMB data. This response is entirely controlled by the expansion history prior to recombination, and is therefore insensitive to any subsequent evolution of the self--interaction component.


\section{Impact on the growth of structure and \texorpdfstring{$S_8$}{S8}}
\label{subsec:growth_S8}

In this section we study how the transient self--interaction component affects the growth of matter perturbations. Its impact enters through two modifications of the background evolution: a temporary enhancement of the expansion rate, $H(a)>H_{\Lambda{\rm CDM}}(a)$, and a corresponding reduction of the fractional matter density, $\Omega_m(a)$. Both effects act to suppress the linear growth of matter fluctuations.

To quantify this response, we treat the self--interaction fraction $f_{\rm SI}(a)$ as a small perturbation and expand the matter density contrast around the standard $\Lambda$CDM solution,
\begin{equation}
    \delta(a) = \delta_0(a) + \delta_1(a),
    \label{eq:delta_split}
\end{equation}
where $\delta_0(a)$ denotes the usual growing--mode solution in $\Lambda$CDM, which determines the late--time amplitudes of $\sigma_8$ and $S_8$. The correction $\delta_1(a)$ captures the linear response to the temporary deviation from $\Lambda$CDM induced by $f_{\rm SI}(a)$.

\subsection{Growth equation in a general background}

The linear growth of matter fluctuations in an arbitrary homogeneous background is governed by
\begin{equation}
  \delta''(a)
  + \left[\frac{3}{a} + \frac{H'(a)}{H(a)}\right]\delta'(a)
  - \frac{3}{2}\,\frac{\Omega_m(a)}{a^2}\,\delta(a) = 0,
  \label{eq:growth_general}
\end{equation}
where primes denote derivatives with respect to the scale factor $a$.
The first two terms represent Hubble friction, while the last term accounts for the gravitational attraction sourced by matter. Any modification of $H(a)$ or $\Omega_m(a)$ therefore feeds directly into the evolution of $\delta(a)$ (see, e.g., \cite{linder2020limited, ishak2019testing}). Notice that equation~\eqref{eq:growth_general} corresponds to the standard linear growth equation in the Newtonian and sub--horizon regime \cite{ma1995cosmological, dodelson2020modern}.

In the $\Lambda$CDM background, this equation reduces to
\begin{equation}
  \delta_0''(a)
  + \left[\frac{3}{a} + \left(\frac{H'}{H}\right)_{\Lambda}\right]\delta_0'(a)
  - \frac{3}{2}\,\frac{\Omega_m^{\Lambda}(a)}{a^2}\,\delta_0(a)=0,
  \label{eq:growth_LCDM}
\end{equation}
whose growing--mode solution defines $\delta_0(a)$.

\subsection{Background modification from the self--interaction fraction}

To first order in the small fraction $f_{\rm SI}\ll1$, the background quantities are modified according to Eq.~\eqref{eq:hubble} and
\begin{equation}
    \Omega_m(a) \simeq \Omega_m^{\Lambda}(a)\,[1 - f_{\rm SI}(a)].
    \label{eq:Omega_m_with_fSI}
\end{equation}

Substituting these expressions into Eq.~\eqref{eq:growth_general} and expanding the density contrast according to Eq.~\eqref{eq:delta_split} to linear order in $f_{\rm SI}$ yields a forced equation for the correction $\delta_1(a)$,
\begin{align}
\delta_1''
+ \left[\frac{3}{a} + \left(\frac{H'}{H}\right)_\Lambda\right]\delta_1'
- \frac{3}{2}\frac{\Omega_m^\Lambda(a)}{a^2}\,\delta_1
= S(a),
\label{eq:growth_forced}
\end{align}
where the source term encodes both the change in Hubble friction and the reduction of the effective matter density during the self--interaction phase,
\begin{equation}
S(a)
= -\left[
\frac{1}{2}f_{\rm SI}'(a)\,\delta_0'(a)
+ \frac{3}{2}\frac{\Omega_m^\Lambda(a)}{a^2}\,
  f_{\rm SI}(a)\,\delta_0(a)
\right].
\label{eq:source_term}
\end{equation}

The differential operator on the left--hand side of Eq.~\eqref{eq:growth_forced} is identical to that governing the $\Lambda$CDM solution $\delta_0(a)$ and therefore admits the same homogeneous modes: a growing mode and a decaying mode. The latter is irrelevant at late times and is not excited by a brief self--interaction episode when the initial conditions are chosen consistently. Using variation of constants, or equivalently the Green’s function of the $\Lambda$CDM operator (see, e.g., \cite{Dodelson}), the solution for $\delta_1(a)$ can be written as
\begin{equation}
  \delta_1(a)
  = \delta_0(a)\int_0^a G(a,\tilde a)\, S(\tilde a)\,d\tilde a,
  \label{eq:delta1_KS}
\end{equation}
where $G(a,\tilde a)$ is a kernel encoding the $\Lambda$CDM linear response. This expression makes explicit how contributions from different epochs accumulate to suppress the growth factor. A similar kernel--based interpretation has been emphasized in studies of EDE and other modified expansion histories \cite{poulin2019early, hill2020early}.

For convenience, we rewrite this result in terms of the linear growth factor $D(a)=\delta(a)/\delta_{\rm ini}$ and its $\Lambda$CDM counterpart $D_0(a)$,
\begin{equation}
\frac{\Delta D}{D_0}(a)
= \frac{\delta_1(a)}{\delta_0(a)}
= \int_0^a G(a,\tilde a)\, S(\tilde a)\, d\tilde a.
\label{eq:dD_over_D_kernel}
\end{equation}

\subsection{Growth kernel representation}

Inserting the explicit source term, Eq.~\ref{eq:source_term}, and integrating by parts to remove
$f_{\rm SI}'$ yields
\begin{equation}
  \frac{\Delta D}{D_0}(a)
  = -\int_0^a W_{\rm gr}(\tilde a)\,
      f_{\rm SI}(\tilde a)\,d\ln\tilde a,
  \label{eq:dD_over_D_Wgr}
\end{equation}
with the growth kernel
\begin{eqnarray}
  W_{\rm gr}(\tilde a)
  &=& \tilde a
  \bigg[
    \frac{3}{2}\frac{\Omega_m^{\Lambda}(\tilde a)}{\tilde a^{2}}
      G(\tilde a)\delta_0(\tilde a) \nonumber\\
  && - \frac{d}{d\tilde a}
      \left(\frac{1}{2}G(\tilde a)\delta_0'(\tilde a)\right)
  \bigg].
  \label{eq:Wgr_def}
\end{eqnarray}

Here we have defined $G(\tilde a)\equiv G(a_0,\tilde a)$ for notational simplicity. The full dependence on the background cosmology is encoded in the kernel $W_{\rm gr}$, while the self--interaction enters only through the quantity $f_{\rm SI}(a)$.

Identifying the fractional change in $S_8$ with the relative change in the linear growth factor, $\Delta S_8/S_8 \simeq \Delta D/D$, and evaluating Eq.~\eqref{eq:dD_over_D_Wgr} at the present epoch $a_0=1$, we obtain
\begin{equation}
  \frac{\Delta S_8}{S_8}
  \simeq
  -\int_0^{1} W_{\rm gr}(a)\,f_{\rm SI}(a)\,d\ln a.
  \label{eq:DeltaS8_general}
\end{equation}

\subsection{Narrow--bump approximation and analytic estimate}

For the self--interaction scenarios of interest, $f_{\rm SI}(a)$ is well approximated by eq. \eqref{eq:fSI}. Since the growth kernel $W_{\rm gr}(a)$ varies slowly across the bump, we may approximate $W_{\rm gr}(a)\simeq W_{\rm gr}(a_\star)$. Equation~\eqref{eq:DeltaS8_general} then reduces to
\begin{align}
  \frac{\Delta S_8}{S_8}
  &\simeq 
  -W_{\rm gr}(a_\star)
   \int f_{\rm SI}(a)\,d\ln a
\nonumber\\[4pt]
  &= -W_{\rm gr}(a_\star)\,f_{\rm pk}\,\sqrt{2\pi}\,\Delta,
  \label{eq:DeltaS8_bump}
\end{align}
This equation provides an analytic estimate for the $S_8$ supression in terms of the position, amplitude, and width of the self--interaction bump, weighted by the growth kernel.

\begin{figure}[t]
  \centering
  \includegraphics[width=0.95\linewidth]{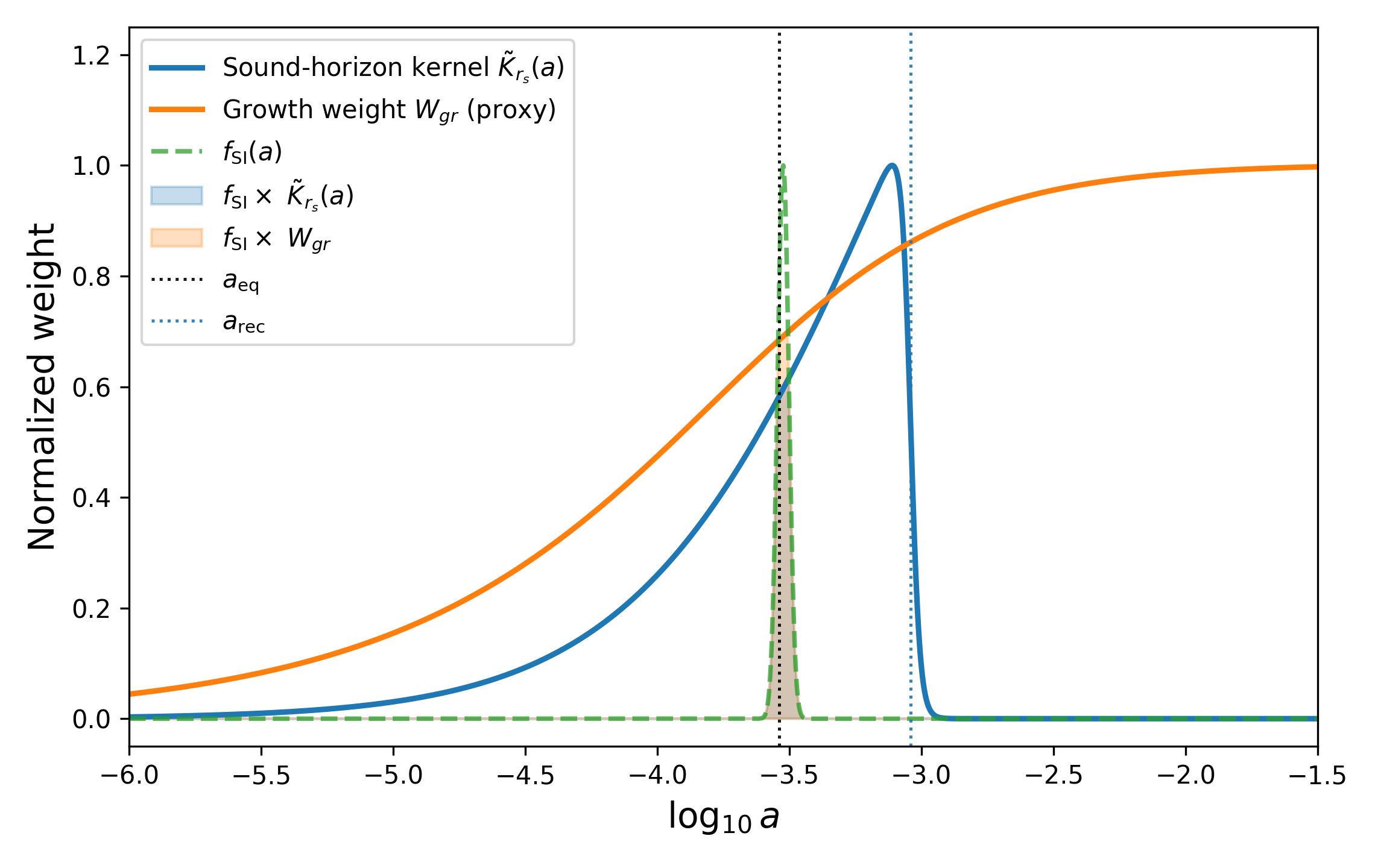}
  \caption{Sensitivity kernels for the sound horizon and linear growth. The blue curve shows the sound–horizon sensitivity, which peaks before recombination and vanishes afterwards, while the orange curve illustrates the cumulative sensitivity of linear growth, becoming effective after matter–radiation equality. The dashed green curve represents a localized self–interaction fraction $f_{\text{SI}}(a)$ centered at $a_{\star}$. Vertical lines mark equality $a_{\text{eq}}$ and recombination $a_{\text{rec}}$. Shaded regions indicate the portions of the self–interaction phase that contribute to the reduction of the sound horizon (blue) and to the suppression of structure growth (orange).}
  \label{fig:kernels}
\end{figure}

Figure~\ref{fig:kernels} illustrates how a single transient self--interaction episode can affect different cosmological observables depending on the epoch at which it occurs. The dashed green curve shows the localized self--interaction fraction $f_{\rm SI}(a)$, which induces a temporary enhancement of the expansion rate. The blue curve represents the sound--horizon kernel $\tilde K_{r_s}(a)$, Eq.~\eqref{eq:sound_kernel}, which peaks prior to recombination and weights the contribution of the expansion history to the acoustic scale. By contrast, the orange curve shows a proxy for the growth kernel $W_{\rm gr}(a)$, whose support lies predominantly after matter--radiation equality and governs the suppression of structure growth. Although the background modification is identical, its observable impact depends on how $f_{\rm SI}(a)$ overlaps with each kernel:  at early times it primarily reduces the sound horizon, while at later times it suppresses the growth of structure.

\section{Combined response of \texorpdfstring{$H_0$}{H0} and \texorpdfstring{$S_8$}{S8}}
\label{sec:combined}

The previous sections have shown that a transient quartic phase of SI--ULDM induces correlated responses in the sound horizon and in the late--time growth of matter perturbations. Although both effects originate from the same self--interaction fraction $f_{\rm SI}(a)$, they probe different epochs of cosmic history and therefore exhibit distinct temporal sensitivities. As a result, a single physical modification does not translate into a one--to--one mapping between the two observables.

\begin{figure*}
    \centering
    \includegraphics[width=0.48\linewidth]{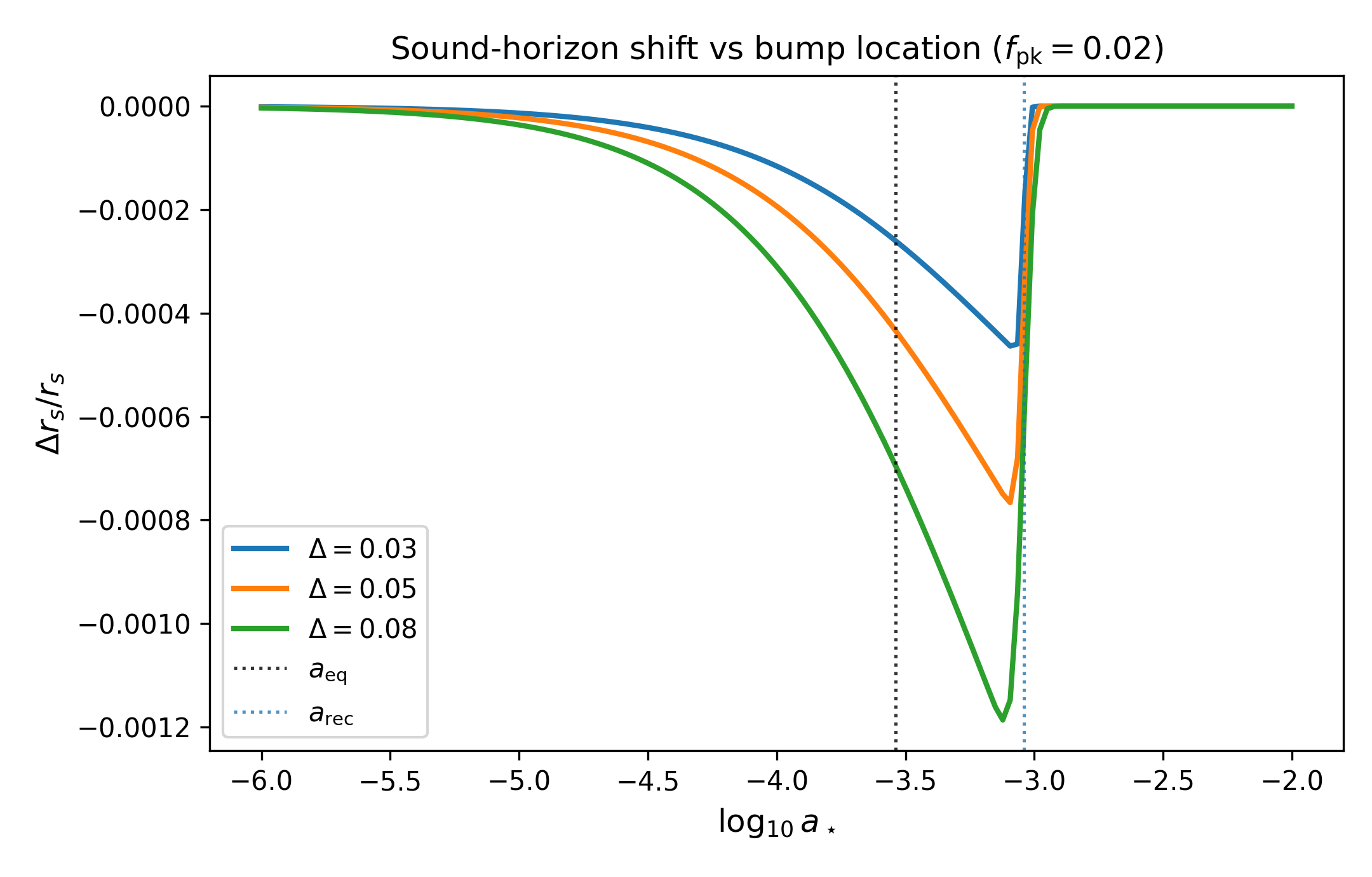}
    \includegraphics[width=0.48\linewidth]{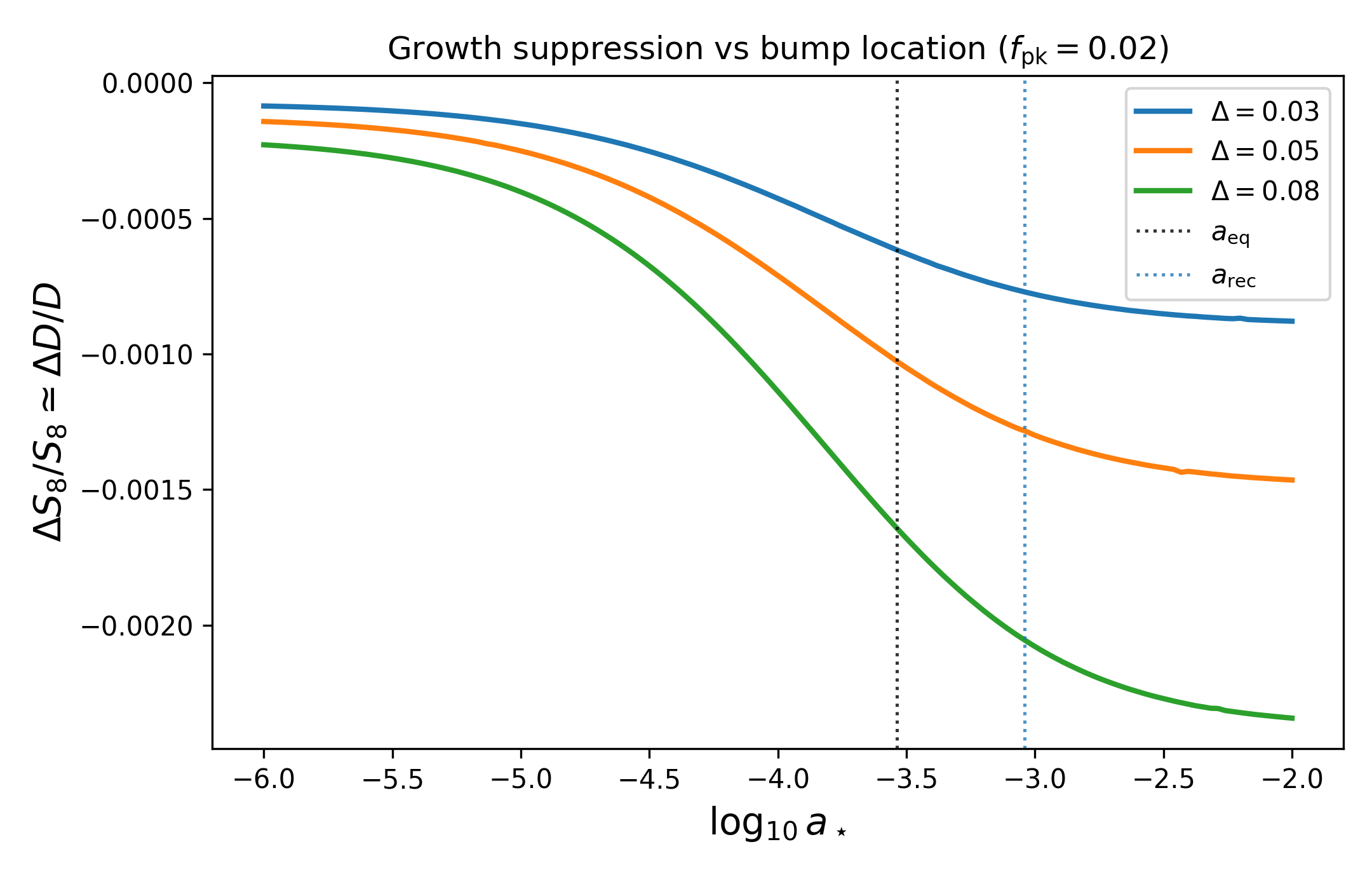}
    \caption{Impact of a transient self–interaction phase on early- and late-time cosmological observables as a function of its characteristic epoch.
\textbf{Left}: fractional shift of the sound horizon at recombination, $\Delta r_s/r_s$, as a function of the bump location $a_\star$, for fixed peak amplitude $f_{\rm pk}=0.02$ and different widths $\Delta$. The effect is maximal when the self–interaction occurs close to matter–radiation equality and rapidly vanishes for transitions taking place well before or after recombination. \textbf{Right}: corresponding fractional suppression of the linear growth amplitude, $\Delta S_8/S_8\simeq\Delta D/D$, obtained by solving the linear growth equation. In contrast to the sound horizon, the impact on structure growth remains significant even when the self–interaction occurs after recombination, reflecting the cumulative nature of growth. Vertical lines indicate matter–radiation equality ($a_{\rm eq}$) and recombination ($a_{\rm rec}$).}
    \label{fig:placeholder}
\end{figure*}

Figure~\ref{fig:placeholder} illustrates this behaviour explicitly. The left panel shows the fractional shift of the sound horizon, $\Delta r_s/r_s$, as a function of the characteristic scale factor $a_\star$ at which the self--interaction bump peaks. Because acoustic oscillations cease at recombination, only the portion of $f_{\rm SI}(a)$ occurring prior to $a_{\rm rec}$ contributes to the sound horizon. Consequently, the impact on $r_s$ is maximal when the self--interaction overlaps with the epoch around matter--radiation equality, and rapidly diminishes for transitions occurring well before or after recombination, in agreement with the structure of the kernel $\tilde K(a)$ discussed in Sec.~\ref{sec:kernel_sound}.

The right panel shows the corresponding fractional suppression of the late--time clustering amplitude, $\Delta S_8/S_8 \simeq \Delta D/D$. In contrast to the sound horizon, the growth of structure remains sensitive to background modifications long after recombination. As a result, the suppression of $S_8$ extends to significantly later values of $a_\star$, reflecting the cumulative nature of gravitational growth during the matter--dominated era.

\subsection{Analytic relation between $\Delta H_0$ and $\Delta S_8$}

From the sound--horizon analysis we obtained Eq.~\eqref{eq:delta_rs}, which, through Eq.~\eqref{eq:delta_h0}, translates into a shift in the CMB--inferred Hubble constant. Combining both expressions yields
\begin{equation}
\frac{\Delta H_0}{H_0}
\simeq
-\frac{1}{2}\,\eta\,f_{\rm pk}\,\mathcal{W}_\star.
\label{eq:dH0_fpk}
\end{equation}
This relation can be inverted to express the peak amplitude of the self--interaction bump in terms of a given shift in $H_0$,
\begin{equation}
f_{\rm pk}
\simeq
-\frac{2}{\eta\,\mathcal{W}_\star}
\,\frac{\Delta H_0}{H_0}.
\label{eq:fpk_from_H0}
\end{equation}

On the other hand, the growth analysis in Sec.~\ref{subsec:growth_S8} yields an analytic expression for the suppression of $S_8$ in the narrow--bump approximation,
Eq.~\eqref{eq:DeltaS8_bump}. Substituting Eq.~\eqref{eq:fpk_from_H0} into this result leads to a direct relation between the shifts in $S_8$ and $H_0$,
\begin{equation}
\frac{\Delta S_8}{S_8}
\simeq
\frac{2\sqrt{2\pi}\,\Delta}{\eta}\,
\frac{W_{\rm gr}(a_\star)}{\mathcal{W}_\star}\,
\frac{\Delta H_0}{H_0}.
\label{eq:S8_H0_relation}
\end{equation}

Equation~\eqref{eq:S8_H0_relation} provides an analytic description of the correlated response of $H_0$ and $S_8$ to a transient self--interaction phase. While both observables depend on the same parameters $(f_{\rm pk},\Delta,a_\star)$, their response is weighted by kernels with different temporal support. The sound horizon is primarily sensitive to modifications near recombination, encoded in the weight $\mathcal{W}_\star$, whereas the growth kernel $W_{\rm gr}(a)$ peaks around matter--radiation equality and during the onset of matter domination.

As a result, the relation between $\Delta H_0$ and $\Delta S_8$ is not universal, but depends on the temporal localization of the self--interaction episode. For bumps centered near matter--radiation equality, $\mathcal{W}_\star>0$ and $W_{\rm gr}(a_\star)>0$, while $\eta<0$, implying that an increase in the inferred value of $H_0$ is generically accompanied by a suppression of $S_8$.

The purpose of this analysis is not to fully resolve the current discrepancies, but to isolate and characterize the structural response induced by a single transient modification of the early--time expansion history within a controlled perturbative framework.

\begin{figure}
    \centering
    \includegraphics[width=\linewidth]{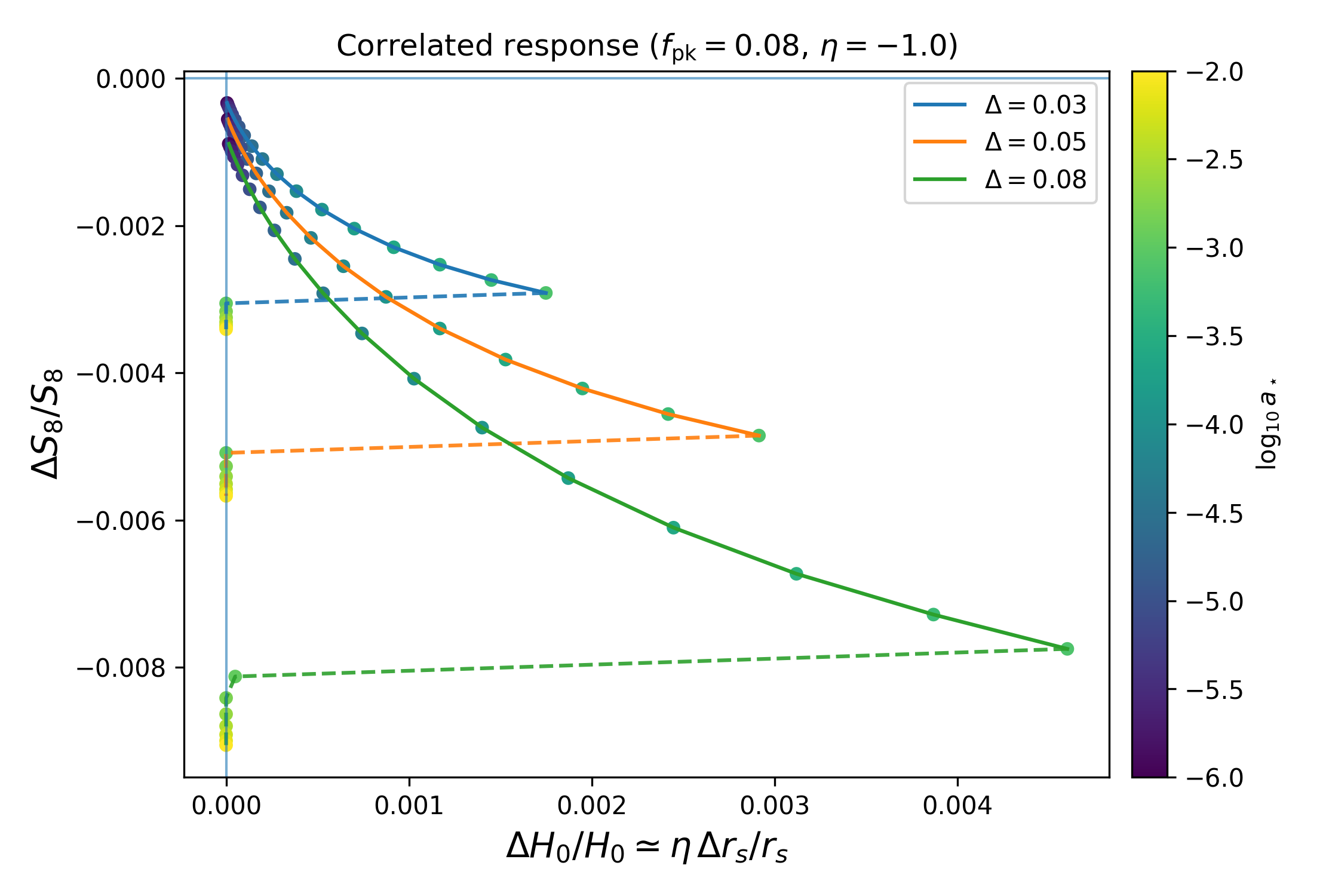}
    \caption{
Correlated response of the Hubble constant and the amplitude of matter fluctuations induced by a transient self--interaction phase. The figure shows the parametric trajectories in the $(\Delta H_0/H_0,\Delta S_8/S_8)$ plane obtained by varying the location of the self--interaction bump, $a_\star$, for fixed peak amplitude $f_{\rm pk}=0.08$ and different widths $\Delta$ (solid colors). The color of the markers encodes the value of $\log_{10} a_\star$. Solid lines correspond to self--interaction episodes occurring before recombination, for which both the sound horizon and the growth of structure are affected. The transition to dashed lines marks the regime in which the bump moves past recombination: beyond this epoch, further changes in $a_\star$ no longer modify the sound horizon, leading to a saturation of $\Delta H_0$, while the suppression of structure growth continues to accumulate. The resulting ``kink'' in the trajectories reflects this change of temporal sensitivity and illustrates that the relation between $\Delta H_0$ and $\Delta S_8$ is not one--to--one, but depends on the epoch at which the transient self--interaction takes place.}
\label{fig:fig3}

    \label{fig:s8_h0}
\end{figure}

Figure~\ref{fig:fig3} shows the parametric trajectory traced in the $(\Delta H_0/H_0,\Delta S_8/S_8)$ plane as the location of the self--interaction bump is varied. The structure of the trajectories reflects the fact that self--interaction episodes occurring before recombination primarily affect $H_0$, while those taking place at later times mainly suppress $S_8$, smoothly interpolating between these two regimes.

For the amplitudes shown, the induced shifts remain modest, as expected in the perturbative regime considered here. This emphasizes that the significance of the mechanism lies in its predictive correlation and temporal structure, rather than in fine--tuned parameter choices.

\section{Conclusions}
\label{sec:conclusions}

In this work we have investigated the cosmological impact of a transient quartic phase of self--interacting ultra--light dark matter (SI--ULDM), focusing on its simultaneous effects on the sound horizon at recombination and on the late--time growth of structure. The key feature of this scenario is that the self--interaction energy density behaves radiation--like at early times and matter--like at late times, producing a localized modification of the expansion history around a characteristic epoch $a_\star$.

We have shown that this single physical ingredient induces correlated shifts in the Hubble constant $H_0$ and in the clustering parameter $S_8$, but through distinct physical mechanisms. The sound horizon is sensitive only to the portion of the self--interaction phase occurring prior to recombination, leading to a reduction of $r_s$ and hence to a larger CMB--inferred value of $H_0$. By contrast, the growth of matter perturbations responds cumulatively to the modified background, allowing the same transient episode to suppress $S_8$ even when it occurs partly or entirely after recombination.

Using a perturbative and analytically controlled framework, we derived simple expressions that relate the shifts in $H_0$ and $S_8$ directly to the amplitude, width, and timing of the self--interaction bump $f_{\rm SI}(a)$. In the narrow--bump limit, this leads to an explicit analytic relation between $\Delta H_0$ and $\Delta S_8$, demonstrating that a positive shift in $H_0$ is generically accompanied by a negative shift in $S_8$ for transitions occurring near matter--radiation equality. This correlated signature is a robust prediction of the SI--ULDM scenario and arises largely independently of the microscopic details of the scalar field, which enter only through the effective parameters describing $f_{\rm SI}(a)$.

Our results highlight a qualitative difference between transient early--time modifications of the expansion history and purely late--time new physics. In particular, the SI--ULDM framework provides a unified physical mechanism that connects the responses of $H_0$ and $S_8$, while predicting a characteristic pattern of correlation that can be tested against cosmological data.

The analysis presented here is by design analytic and phenomenological, aimed at isolating the underlying physical mechanisms. A full quantitative assessment will require implementing the SI--ULDM background evolution in a Boltzmann solver such as \textsc{CLASS} and considering the model with CMB, BAO, and large--scale structure data. Extensions including perturbations of the scalar field itself, as well as possible interactions with baryons or neutrinos, are left for future work.

\section*{Acknowledgements}
JNLS acknowledges the support by the European Union and the Czech Ministry of Education, Youth and Sports (Project: MSCA Fellowships CZ FZU III -CZ$.02.01.01/00/22\_010/0008598$)

\bibliographystyle{unsrt}
\bibliography{references}

\end{document}